\documentclass[aoas,MSNbibl,nameyear,dvips]{arximspdf}

\doi{10.1214/14-AOAS728R} 
\volume{8}
\issue{4}
\pubyear{2014}
\firstpage{1961}
\lastpage{1965}
\docsubty{FLA}
\referstodoi{10.1214/14-AOAS728}


\begin{document}
\begin{frontmatter}

\title{Rejoinder: ``Spatial accessibility of pediatric primary
healthcare: Measurement and
inference''}
\runtitle{Rejoinder}

\begin{aug}
\author[A]{\fnms{Mallory}~\snm{Nobles}\ead[label=e1]{mallory.nobles@gmail.com}},
\author[A]{\fnms{Nicoleta}~\snm{Serban}\corref{}\ead[label=e2]{nserban@isye.gatech.edu}}
\and
\author[A]{\fnms{Julie}~\snm{Swann}\ead[label=e3]{swann@isye.gatech.edu}}
\runauthor{M. Nobles, N. Serban and J. Swann}
\affiliation{Georgia Institute of Technology}
\address[A]{H. Milton Stewart School of Industrial\\
\quad and Systems Engineering\\
Georgia Institute of Technology
755 Ferst Dr. NW\\
Atlanta, Georgia 30332\\
USA\\
\printead{e1}\\
\phantom{E-mail:\ }\printead*{e2}\\
\phantom{E-mail:\ }\printead*{e3}}
%

\end{aug}

\received{\smonth{8} \syear{2014}}


\end{frontmatter}

We first would like to thank all discussants for their thoughtful
comments. We appreciate the additional insights regarding our findings
and the suggestions on future directions relevant to the estimation of
and inference on healthcare access. In our rejoinder, we emphasize
three discussion threads addressing challenges and limitations of the
proposed methodology, and addressing further considerations in the
interpretation of our models with implications in informed decision making.

\emph{Local estimates and targeted interventions.} In recent years, the
fields of healthcare services research and health policy have
acknowledged and stressed the significance of applying operations
research methodology to understand and manage the complexity of
healthcare [\citet{RouSer14}]. The methodology can emphasize
impact and improvements at different levels---individuals,
communities, processes, providers, organizations and/or the entire
ecosystem of care. Depending on what actions are targeted in improving
healthcare delivery, one may assess individual-level improvements
(e.g., personalized medicine) or system-wide effects (e.g., health
policy), for example. Our study primarily emphasizes health policy,
while not losing sight of its potential unintended consequences at the
community level. In policy decision making, low geographic granularity
inferences, block or census tracts, are desirable over coarser
geographic aggregations such as county because they will capture the
diversity of populations in need of better care and the diversity of
environments of care delivery.

Making inferences at low geographic granularity, specifically local
assessments of systematic disparities, is not a new research direction
in the medical literature, although one will find novel modeling
contributions in recent years with the advancement of computational
approaches in the area of geographic economics and environmental studies.

An immediate application of local disparity inferences is suggesting
targeted actions to improve various aspects of healthcare. We divide
such actions into policy and network interventions. For example, in the
discussion paper, we primarily focus on \textit{policy interventions},
which commonly involve designing, implementing or translating a health
policy. On the other hand, \textit{network interventions} refer to actions
that involve altering an existing network of care, including location
and allocation of sites. Designing and evaluating such interventions
require understanding healthcare access and its impact on health
outcomes at the community level along with advanced mathematical
modeling, including location-allocation models [\citet{DasDea04}]
and discrete simulations [\citet{JacHalSwi06}], for optimal
allocation of resources while achieving high levels of equity and
quality of care.

Because we find that the policy interventions explored in our study
have limited impact in improving access, the discussants have suggested
two possible network interventions. One intervention is to consider new
service locations for primary care to reduce geographic inequities
and/or reduce the disparities between the Medicaid and non-Medicaid
population. A second intervention is to incentivize providers to
consider satellite care opportunities in more remote areas than
otherwise preferred. The two interventions, however, can be considered
together by designing alternative care opportunities including mobile
clinics and telemedicine, which can use the expertise of physicians in
densely served areas to provide care to patients in underserved and
unserved areas. The optimization and statistical models developed in
our study can be used to derive optimal assignments between providers
and areas in need for care under such alternative care approaches. The
optimal allocation of limited resources can be based on equity and
effectiveness objectives jointly [\citet{GraCarVer14}]. We
are examining each of these interventions in our ongoing research.

\emph{Data uncertainty.} A third thread of comments refers to data
uncertainties in informing the optimization model for measuring access.
We reiterate their importance by providing additional insights and discussions.

In our study, one emphasis is to develop optimization-based measurement
models for estimating spatial access because we recognize that the
existing approaches provide inaccurate estimates due to deficiencies of
the models employed, particularly because of their lack of
incorporating knowledge about the trade-offs and realistic constraints
in the care system. However, errors in the spatial access measures are
also largely driven by uncertainties as a potential deficiency that is
due to lack of knowledge; in statistical terms, these are called
estimation errors. All existing spatial access measures, regardless of
whether they are derived using simple approaches or obtained using more
sophisticated models, are point estimates derived using uncertain data,
and thus they are limited in terms of the ability of making inferences
on spatial access. To the best of our knowledge, this important aspect
is overlooked in the existing literature of spatial access.

Particularly, the optimization assignment model used to measure spatial
access relies on a set of parameters that are assumed fixed and known,
including the maximum willingness to travel for patients with or
without a car, and Medicaid participation along with the maximum
caseload devoted to Medicaid patients of pediatricians in the network
of care. These parameters are specified based on either subjective care
delivery recommendations or sources of data observed with uncertainty.
Moreover, there are also uncertainties in the supply (providers) and
demand (patients) data, due to irregularities in the reporting of
providers and their taxonomy (supply uncertainty) and because of the
estimation error in population counts derived by the Census Bureau
(demand uncertainty). Last, the travel distance from a community to a
provider will vary depending on where patients in the respective
community live and depending on the variations in alternative travel
routes. Such uncertainties may also be different in rural areas than in
urban areas. Thus, the spatial access measures derived from the
optimization models are point estimates, with an estimation error
varying across space and compounding many sources of uncertainty.

We have made an attempt to reduce the uncertainty due to physician
participation in the Medicaid insurance program and we have studied the
sensitivity of our model to small variations in the model parameters;
however, we have also overlooked the error quantification due to other
sources of uncertainty. The discussants have rightly so stressed the
importance of considering quantification of the estimation error in a
principled manner, as it is key in making inferences on local
disparities in spatial access.

Uncertainty quantification has broadly been studied in experimental
design and simulations [\citet{BarNelXie14}; Smith (\citeyear{Smi14});
\citet{Sny06}]. But in studies relying on deterministic optimization
models, the presence of the estimation error due to data uncertainty is
rarely acknowledged. In such models, the~effect of uncertainties of the
model parameters on the estimates derived from the optimization model
is often evaluated using sensitivity analysis and running the model
with different input values; however, quantification of the overall
estimation error is overlooked. Thus, there is an immediate need for
developing methodology to address the need of quantifying the
estimation error not only for spatial access measures, but also for
decision variables and solutions derived from deterministic
optimization models more generally that rely on uncertain data and
uncertain parameters.

\emph{Interpretation of the results.} The discussants suggested several
interesting directions for interpretation of the results of our model.
First, as mentioned in the paper and highlighted in the discussion,
there are strong correlations between dimensions of access to
healthcare. Indeed, the state-wide spatial correlation between average
distance traveled and percent covered is $-0.92.$ The correlation between
distance and congestion is 0.51 and between congestion and coverage is
$-0.45$. In some regions and for some sub-populations, these correlations
will be stronger. Understanding the small scale correlations is
especially important since they indicate where improving access for
some access dimension or population groups may have unintended
consequences. The type or magnitude of correlation can also be useful
in thinking about the type of intervention appropriate in a particular area.

Each discussant suggested further analysis of the striking differences
between access in urban and rural areas of Georgia. In our optimization
model specifications, we chose to set some parameters constant
throughout the state. For example, the parameter for maximum distance
families are willing to travel is specified to be consistent with
policy makers' goal for accessible primary care. If the research
objective is to identify areas with access below a certain threshold,
or locations which experience inequities, this modeling approach may be
sufficient. However, as the discussants correctly point out, this
parameter value will surely vary by location since many families living
in rural areas will tolerate driving further than those living in urban
areas to reach medical services. Therefore, the current approach might
underestimate access in rural areas relative to urban areas and fail to
identify locations that face the strongest spatial inequities in
access. Allowing this parameter and other inputs to the optimization
model to vary across space would result in more accurate measures of
access to healthcare in vulnerable regions.

Additional investigation into the local effects of potential policies
is another important next step for this research. Facility location
methodology could be used to identify regions where a policy would have
the largest effect on the local population. It might also be useful to
understand the degree to which a policy would need to be implemented
before access improves in a particular rural region. The concentration
of need or demand for healthcare will impact a policy's potential
effectiveness. For example, under a fixed budget, a policy maker may
have to choose between an intervention that slightly improves access
for many patients in an urban area and an intervention that
dramatically improves access for a smaller number of patients in a
rural area. Analysis which allows policy makers to better understand
these types of trade-offs would allow for more informed decision making.

Finally, as one discussant noted, further work could be done to
consider the policy implications of the effects of covariates across
the multiple regression models. For example, we found that when
population density had a nonconstant effect on distance traveled,
segregation levels typically had a significant and constant effect.
However, in models where population density had a constant effect,
segregation levels took on a nonconstant effect. Further exploration of
these types of relationships might help policy makers understand the
complex dynamics of access to healthcare. To our knowledge, little work
has been done to develop methodologies that systematically identify and
effectively convey these types of relationships. Further work in this
area has the potential to offer contributions to this line of work and
many other domains.


%



\printaddresses
\end{document}